\documentclass[preprint,12pt]{elsarticle}

\usepackage{subfigure}
\usepackage{amssymb}

\newtheorem{theorem}{Theorem}[section]

\newtheorem{corollary}{Corollary}[section]

\journal{Elsevier Science}

\begin{document}

\begin{frontmatter}


\title{An End-to-End Stochastic Network Calculus with Effective Bandwidth and Effective Capacity}


\author{Kishore Angrishi}
\address{T-Systems International GmbH\\
20146 Hamburg, Germany\\
kishore.angrishi@t-systems.com}

\begin{abstract}
Network calculus is an elegant theory which uses envelopes to determine the worst-case performance bounds in a network. Statistical network calculus is the probabilistic version of network calculus, which strives to retain the simplicity of envelope approach from network calculus and use the arguments of statistical multiplexing to determine probabilistic performance bounds in a network. The tightness of the determined probabilistic bounds depends on the efficiency of modelling stochastic properties of the arrival traffic and the service available to the traffic at a network node. The notion of effective bandwidth from large deviations theory is a well known statistical descriptor of arrival traffic. Similarly, the notion of effective capacity summarizes the time varying resource availability to the arrival traffic at a network node. The main contribution of this paper is to establish an end-to-end stochastic network calculus with the notions of effective bandwidth and effective capacity which provides efficient end-to-end delay and backlog bounds that grows linearly in the number of nodes ($H$) traversed by the arrival traffic, under the assumption of independence.
\end{abstract}


\end{frontmatter}
\section{Introduction}
\label{sec:intro}
The increasing share of real-time traffic (IPTV, VoIP, Internet Radio, etc.,) over the Internet has motivated the study of Quality of Service (QoS) guarantees in data networks. The key aspect of guaranteeing QoS in a data network is to be able to efficiently model the arrival traffic and the service offered by the network. There exist many network theories which facilitate the modelling of traffic and service in data networks. Network calculus is one of the popular theories in recent times useful for the performance analysis of data networks. Network calculus uses deterministic arrival and service envelopes to bound traffic arrivals and the service offered at the nodes, respectively, to compute the worst case end-to-end performance bounds. However, most of the multimedia traffic observed in the Internet can tolerate some violation in its QoS requirements, and moreover, the statistical multiplexing in data networks smoothens the burstiness of the aggregate arrival traffic with high probability. Therefore the theory of network calculus was extended to the probabilistic domain, especially to benefit from the statistical multiplexing in data networks. The probabilistic version of network calculus is called statistical network calculus \footnote{The terms statistical network calculus, stochastic network calculus and probabilistic network calculus are used interchangeably in the literature} and it strives to retain many of the favourable characteristics of the network calculus, especially the simple envelope approach to derive probabilistic bounds. The main issue with the envelope approach employed in statistical network calculus is that the utilization of statistical multiplexing for network analysis is limited to network ingress, as the stochastic information of the arrival traffic is lost once the statistical arrival envelope is derived at the network ingress and no statistical multiplexing can be considered inside the network \cite{KIVS:2007,fidler:2006}. This leads to not so efficient end-to-end delay and backlog bounds in a feed-forward network. 

Probably the most influential related analytical technique used to model stochastic arrival traffic at a network node is effective bandwidth \cite{kelly:1996} from large deviations theory. Effective bandwidth describes the minimum bandwidth required at a network node to provide an expected QoS for a given traffic. Similarly, the concept of effective capacity \cite{kumar:2001,wu:2003,MMB:2008} from large deviations theory can be used to represent the stochastic service received by an arrival traffic at a network node. The QoS measures at a network node can be expressed using the large deviations theory under many sources limiting regime (infinite sources) in terms of effective capacity and effective bandwidth. In spite of the successes in analyzing the single node case, there has been only limited success at identifying end-to-end measures for the network models using large deviations theory.

The key contribution of this paper is that we develop an end-to-end stochastic network calculus based on the notions of effective bandwidth \cite{kelly:1996} and effective capacity \cite{kumar:2001,wu:2003,MMB:2008} from large deviations theory. Such an extension of end-to-end stochastic network calculus allows the stochastic information about the arrival traffic and the service available to an arrival traffic at a network node to be retained as long as possible using the concept of effective bandwidth and effective capacity, respectively, which in turn enables efficient computation of end-to-end stochastic delay and backlog bounds than in other approaches with statistical envelopes \cite{yuming:2006,florin:2006,li:2007,jiang:2009-1} where the stochastic information about the arrival and the service processes are lost as soon as statistical envelopes are fixed. 

The raison d'\^etre of network calculus is the possibility to model a network of nodes as a single abstract node using a network service envelope. In \cite{boudec:2001}, authors have shown that the end-to-end worst-case performance measures obtained by summing the per-hop results are bounded by ${\cal O}(H^2)$, while the end-to-end bounds obtained using network service envelope scales linearly in  the number of nodes $H$ connected in series. There have been many attempts to achieve similar linear scaling of end-to-end probabilistic performance bounds in statistical network calculus but with limited success. Most of these attempts use network service envelope with conservative envelope definitions \cite{yuming:2006} or with some mathematical extensions like, rate correction factor \cite{florin:2006}, delay threshold, busy period bounds \cite{li:2007}, time-domain extensions \cite{jiang:2009-1}. Most notably, in \cite{florin:2006} authors employed network service envelope with rate correction factor to compute the end-to-end performance measures that scales as ${\cal O}(H \log{H})$. In \cite{fidler:2006}, authors have shown using stochastic network calculus with moment generating functions that, if the arrival traffic and the service offered at each network node are independent of one another, the end-to-end performance bounds can scale linearly, i.e., ${\cal O}(H)$. We direct the interested readers to the corresponding papers and to \cite{li:2007} for an elaborate discussion on what makes statistical network calculus so difficult. 

In the later part of this paper, we show that the end-to-end QoS measures determined using newly developed stochastic network calculus with effective bandwidth and effective capacity also scales linearly under the assumption of existence of effective bandwidth and effective capacity for arrival and service processes, respectively, and the independence of processes. Moreover, the usage of effective bandwidth and effective capacity functions to describe the arrival and service processes, respectively, at the network node will nullified any necessity to use rate correction factor \cite{florin:2006} or delay threshold, busy period bounds \cite{li:2007}, time-domain extensions \cite{jiang:2009-1} in network calculus and allows efficient representation of stochastic processes to compute tighter bounds even for single node case.


The rest of the paper is structured as follows: Section \ref{sec:models} introduces the arrival and service model used in the paper. In Section \ref{sec:SNC}, we derive end-to-end backlog and delay bounds using stochastic network calculus with effective bandwidth and effective capacity functions. In Section \ref{sec:num}, a numerical example using Markov Modulated On-Off traffic is presented for illustration. Brief conclusions are presented in Section \ref{sec:conclusion}.  

\section{Arrival and Service Models}
\label{sec:models}

\begin{figure}
\centering
\includegraphics[scale=0.45]{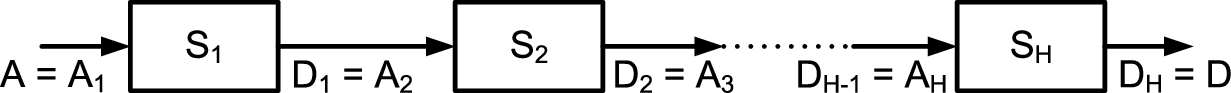}
\caption{ Network of H concatenated nodes}
\label{fig:tandemnet} 
\vspace{-5 mm}
\end{figure}

In this section, we give a brief overview of the arrival and service models employed in this paper. Throughout this paper, we use a discrete time model $t \in \mathbb{N}_0 = \{0, 1, 2, \ldots \}$. We consider a network of $H$ nodes connected in series as shown in Fig. \ref{fig:tandemnet}, with each node in the network having infinite-sized buffer serves the arrival traffic in a work-conserving fashion. Let the arrival and departure processes at a network node $h$ are modelled with bivariate real-valued left-continuous processes $A_h(s,t)$ and $D_h(s,t)$, respectively, which represents the cumulative amount of data seen in the interval $(s, t]$ for any $0 \le s \le t$. Let the service offered at hop $h$ for $h=1, \ldots, H$ is characterized using a bivariate real-valued left-continuous process $S_h(s,t)$, which represents the cumulative amount of data served at the node in the interval $(s,t]$ for any $0 \le s \le t$. To simplify the notation, we denote $A_h(0,t) = A_h(t), D_h(0,t) = D_h(t), S_h(0,t) = S_h(t)$ for any $t \ge 0$. We assume that the network is causal, i.e., $A_h(t) \le D_h(t)$ at any hop $h$ in the network for any $t \ge 0$, and there are no arrivals in the interval $(-\infty, 0]$. For an arrival process $A_h$ at a network node $h$, whose offered service is characterized by a stochastic service process $S_h$, the corresponding departure process $D_h$ satisfies for any fixed sample path and all $t \ge 0$:
\begin{equation}
 A_h\otimes S_h(t) \le D_h(t)
 \label{reffsenv}
\end{equation}
where $\otimes$ denotes the $(min,+)$ convolution of $A_h$ and $S_h$ which is defined as $A_h \otimes S_h(t) = \inf_{0 \le s \le t} \{A_h(0,s) + S_h(s,t)\}$. Any random process $S$ satisfying the above relationship (equation (\ref{reffsenv})) between arrival process and departure process for any fixed sample path is referred to as ``dynamic F-server'' \cite{chang:2000}. 

Let $A=A_1$ be the arrival traffic at the node $1$ or ingress of the network and $D=D_{H}=A_{H+1}$ represent the departure traffic from the node $H$ or egress of the network as shown in Fig. \ref{fig:tandemnet}. The departure traffic $D_{h}$ from the node at hop $h$ becomes the arrival traffic $A_{h+1}$ to the downstream node at hop $h+1$, i.e., $A_{h+1}=D_h$ for all $h=1,\ldots, H$. In \cite{chang:2000,fidler:2006}, authors show that the stochastic network service process $S_{net}$ describing the service offered in network of $H$ nodes connected in series, with stochastic service process $S_h$ for $h=1, \ldots, H$ characterizing the corresponding service offered at hop $h$, for any fixed sample path is given by 
\begin{equation}
S_{net} = S_1 \otimes S_2 \otimes \cdots \otimes S_H
\label{netreffsenv} 
\end{equation}

To derive probabilistic performance measures using the stochastic arrival and service processes, one needs non-random functions characterizing the arrival and service processes. In \cite{fidler:2006}, authors used moment generating function of the arrival traffic and conjugate moment generating function of the service process to derive the performance measures in a network. Here we adopt the popular notions of effective bandwidth ($\alpha_h$) \cite{kelly:1996} and effective capacity ($\beta_h$) \cite{kumar:2001,wu:2003,MMB:2008} from large deviations theory to describe the stochastic arrival process and the service process characterizing the service offered at a network node $h$, respectively. The effective bandwidth of an arrival traffic $A_h$ from \cite{kelly:1996}, for all $ \theta, t > 0$, is given as
\begin{equation}
\alpha_h(\theta,t) = \frac{1}{\theta t} \log{E\left[ e^{\theta A_h(t)}\right]}  
\label{eb}
\end{equation}
Similarly, the effective capacity function of a stochastic service process $S_h$ at a node $h$ from \cite{MMB:2008}, for all $\theta, t > 0$, is defined as
\begin{equation}
{\beta_h}(\theta,t) = - \frac{1}{\theta t} \log{E\left[ e^{-\theta  S_h(t)}\right]}
	 	\label{ec}
\end{equation} 

\section{Stochastic Network Calculus with Effective Bandwidth and Effective Capacity}
\label{sec:SNC}
In this section we apply the arrival and service models from Section \ref{sec:models} to derive end-to-end performance bounds using stochastic network calculus with effective bandwidth and effective capacity. The fundamental difference between the statistical network calculus and its deterministic counterpart is that the performance bounds are expressed as probabilistic tail bounds, i.e., the derived bounds are violated with some probability. The end-to-end backlog $B$ and delay $W$ processes at time $t$ in network of $H$ nodes connected in series as shown in Fig. \ref{fig:tandemnet} are given by $B(t) = A(t) - D(t)$ and $W(t) = \inf{\{d \ge 0 : A(t - d) \le D(t)\}}$, respectively. In the following we derive the probabilistic bound on the end-to-end backlog and delay using stochastic network calculus with effective bandwidth and effective capacity functions. It should be noted that no assumptions on the independence of arrival and service processes were made.
\begin{theorem}
\label{theorem:pbr}
Let $A$ be the arrival traffic to a network of $H$ nodes connected in series with effective bandwidth function $\alpha$ and $D$ be the departure traffic from the network. Assume $S_h$ for $h=1, \ldots, H$ be the stochastic service process at each hop in a network of $H$ nodes with their corresponding effective capacity $\beta_h$. Then we have the following probabilistic bounds.
\begin{enumerate}
	\item Backlog bound : The probabilistic bound on the backlog in a network is given, for all $t \ge 0$, by
				\begin{eqnarray}
				P\left\{B(t) > x \right\} &\le& \inf_{\theta > 0} \sum_{u_H=0}^{t} \cdots \sum_{u_2=0}^{u_3}\sum_{u_1=0}^{u_2} e^{\frac{\theta}{H+1} \left((t-u_1)\alpha(\theta,t-u_1) \right)} \cdot \nonumber\\
				& &  e^{-\frac{\theta}{H+1} \left((u_2-u_1)\beta_1(\theta,u_2-u_1)+ \cdots +(t-u_H)\beta_H(\theta,t-u_H) + x \right)}	\label{backlogr} 
				\end{eqnarray}	
	\item Delay bound : The probabilistic bound on the delay in a network is given, for all $t \ge 0$, by
				\begin{eqnarray}
				P\left\{W(t) > d \right\}  &\le& \inf_{\theta > 0} \sum_{u_H=d}^{t} \cdots \sum_{u_2=d}^{u_3}\sum_{u_1=d}^{u_2} e^{\frac{\theta}{H+1} \left((t-d-u_1)\alpha(\theta,t-d-u_1) \right)} \cdot \nonumber\\
				& & \ \ \ e^{-\frac{\theta}{H+1} \left((u_2-u_1)\beta_1(\theta,u_2-u_1)+ \cdots +(t-u_H)\beta_H(\theta,t-u_H)\right)}	\label{delayr}
				\end{eqnarray}
\end{enumerate}
\end{theorem}
\textbf{\textit{Proof:}}
We now prove the probabilistic bound on backlog $B$. For any $t \ge 0$, we have 
\begin{eqnarray*}
\lefteqn{P\left\{ B(t) > x \right\} = P\left\{ A(t) - D(t) > x \right\} \le P\left\{ A(t) - A\otimes S_{net}(t) > x \right\}}\\
&=& P\left\{ A(t) - A\otimes S_{1}\otimes \cdots \otimes S_{H}(t) > x \right\}\\
&=& P\left\{  \sup_{0 \le u_1 \le t}\{A(t) - A(0,u_1) - S_{1}\otimes \cdots \otimes S_{H}(u_1,t)\} > x \right\}\\
&=&  P\left\{ \sup_{0 \le u_1 \le u_2 \le u_3 \le \cdots \le u_H \le t} \left\{ A(u_1,t) - \{ S_1(u_1,u_2) \right.\right.\\
&& \ \ \ \ \ \ \ \ \ \ \ \ \ \ \ \ \ \ \ \ \ \ \ \ \ \ \ \ \ \ \ \ \ \ \ \ \ \ \ \ \ \ \left.\left. +  S_2(u_2,u_3) + \cdots + S_H(u_H,t) \} \right\} > x \right\}\\
&\le&  E\left[ e^{\Theta \sup_{0 \le u_1 \le u_2 \le u_3 \le \cdots \le u_H \le t} \left\{ A(u_1,t) - S_1(u_1, u_2) - S_2(u_2, u_3) - \cdots - S_H(u_H, t) \right\}} \right] e^{-\Theta x}\\
&\le& \sum_{u_H=0}^{t} \cdots \sum_{u_2=0}^{u_3}\sum_{u_1=0}^{u_2} E\left[ e^{\Theta \left\{ A(u_1,t) - S_1(u_1, u_2) - S_2(u_2, u_3) - \cdots - S_H(u_H, t) \right\}} \right] e^{-\Theta x} \\
&\le& \sum_{u_H=0}^{t} \cdots \sum_{u_2=0}^{u_3}\sum_{u_1=0}^{u_2} E\left[ e^{\Theta (H+1) A(u_1,t)} \right]^{\frac{1}{H+1}} E\left[ e^{- \Theta (H+1) S_1(u_1, u_2)}\right]^{\frac{1}{H+1}} \\
&& \ \ \ \ \ \ \ \ \ \ \ \ \ \ \ \ \ \ \ E\left[ e^{- \Theta (H+1) S_2(u_2, u_3)}\right]^{\frac{1}{H+1}} \cdots E\left[ e^{- \Theta (H+1) S_H(u_H, t)}\right]^{\frac{1}{H+1}} e^{-\Theta x} \\
&=& \sum_{u_H=0}^{t} \cdots \sum_{u_2=0}^{u_3}\sum_{u_1=0}^{u_2} e^{\frac{\theta}{H+1} \left((t-u_1)\alpha(\theta,t-u_1) \right)} \cdot \\
&& \ \ \ \ \ \ \ \ \ \ \ \ \ \ \ e^{-\frac{\theta}{H+1} \left((u_2-u_1)\beta_1(\theta,u_2-u_1)+ (u_3-u_2)\beta_2(\theta,u_3-u_2) \cdots +(t-u_H)\beta_H(\theta,t-u_H) + x \right)}
\end{eqnarray*}
The proof of the probabilistic bound on delay $W$ follows the similar steps. For any $t \ge 0$, we have
\begin{eqnarray*}
\lefteqn{P\left\{ W(t) > d \right\} = P\left\{ A(t-d) - D(t) > 0 \right\} \le P\left\{ A(t-d) - A\otimes S_{net}(t) > 0 \right\} }\\
&=& P\left\{ A(t-d) - A\otimes S_{1}\otimes \cdots \otimes S_{H}(t) > 0 \right\}\\
&=& P\left\{  \sup_{d \le u_1 \le t}\{A(t-d) - A(0,u_1) - S_{1}\otimes \cdots \otimes S_{H}(u_1,t)\} > 0 \right\}\\
&=&  P\left\{ \sup_{d \le u_1 \le u_2 \le u_3 \le \cdots \le u_H \le t} \left\{ A(u_1,t-d) - \{ S_1(u_1,u_2) \right.\right.\\
&& \ \ \ \ \ \ \ \ \ \ \ \ \ \ \ \ \ \ \ \ \ \ \ \ \ \ \ \ \ \ \ \ \ \ \ \ \ \ \ \ \ \ \left.\left. +  S_2(u_2,u_3) + \cdots + S_H(u_H,t) \} \right\} > 0 \right\}\\
&\le&  E\left[e^{\Theta \sup_{d \le u_1 \le u_2 \le u_3 \le \cdots \le u_H \le t} \left\{ A(u_1,t-d) - S_1(u_1, u_2) - S_2(u_2, u_3) - \cdots - S_H(u_H, t)\right\}}\right]\\
&\le& \sum_{u_H=d}^{t} \cdots \sum_{u_2=d}^{u_3}\sum_{u_1=d}^{u_2} E\left[e^{\Theta \left\{ A(u_1,t-d) - S_1(u_1, u_2) - S_2(u_2, u_3) - \cdots - S_H(u_H, t)\right\}}\right]\\ \\
&\le& \sum_{u_H=d}^{t} \cdots \sum_{u_2=d}^{u_3}\sum_{u_1=d}^{u_2} E\left[ e^{\Theta (H+1) A(u_1,t-d)} \right]^{\frac{1}{H+1}} E\left[ e^{- \Theta (H+1) S_1(u_1, u_2)}\right]^{\frac{1}{H+1}} \\
&& \ \ \ \ \ \ \ \ \ \ \ \ \ \ \ \ \ \ \ \ \ \ \ \ \ E\left[ e^{- \Theta (H+1) S_2(u_2, u_3)}\right]^{\frac{1}{H+1}} \cdots E\left[ e^{- \Theta (H+1) S_H(u_H, t)}\right]^{\frac{1}{H+1}}\\
&=& \sum_{u_H=d}^{t} \cdots \sum_{u_2=d}^{u_3}\sum_{u_1=d}^{u_2} e^{\frac{\theta}{H+1} \left((t-d-u_1)\alpha(\theta,t-d-u_1) \right)} \cdot \\
&& \ \ \ \ \ \ \ \ \ \ \ \ \ \ \ \ \ \ e^{-\frac{\theta}{H+1} \left((u_2-u_1)\beta_1(\theta,u_2-u_1)+ (u_3-u_2)\beta_2(\theta,u_3-u_2) \cdots +(t-u_H)\beta_H(\theta,t-u_H)\right)}
\end{eqnarray*}
The inequalities in the respective parts of the proof for probabilistic backlog and delay bounds follow the similar reasons. The first inequality is from the property of stochastic network service process (equation (\ref{reffsenv})). The second inequality is due to the application of Boole's inequality. The third and fourth inequalities are from Chernoff's bound\footnote{For random variable $X$ and $x, \theta \ge 0$, $P\{X > x\} \le E[e^{\theta X}]e^{-\theta x}$.} and H\"older's inequality\footnote{For random variables $X,Y$ and $a, b > 0$ with $1/a + 1/b = 1$, $E[XY] \le E[X^a]^\frac{1}{a}E[Y^b]^\frac{1}{b}$.}, respectively.  The final step is obtained by setting $\theta = (H+1)\Theta$ and from the definition of effective bandwidth and effective capacity. Minimizing the expression over $\theta$ proves our claim on probabilistic backlog and delay bound.$\blacksquare$\\ 
Since the use of H\"older's inequality can be avoided for the independent random variables\footnote{For any two independent random variables $X,Y$,  $E[XY] = E[X]E[Y]$}, the performance bounds from Theorem \ref{theorem:pbr} can be further improved if the arrival traffic process $A$ and the stochastic service process $S_h$ at each hop $h$ for $h = 1, 2, \ldots, H$ are statistically independent of one another, and is given by the following corollary.
\begin{corollary}
\label{corollary:pbrc}
If the arrival traffic $A$ with effective bandwidth $\alpha$ and the stochastic service process $S_h$ with effective capacity $\beta_h$ for $h=1, \ldots, H$ are independent of one another. Then we have the following probabilistic bounds for the network shown in Fig. \ref{fig:tandemnet}.
\begin{enumerate}
	\item Backlog bound : The probabilistic bound on the backlog in a network is given, for all $t \ge 0$, by
				\begin{eqnarray}
				P\left\{B(t) > x \right\} &\le& \inf_{\theta > 0} \sum_{u_H=0}^{t} \cdots \sum_{u_2=0}^{u_3}\sum_{u_1=0}^{u_2} e^{\theta \left((t-u_1)\alpha(\theta,t-u_1) \right)} \cdot \nonumber\\
				& &  e^{-\theta \left((u_2-u_1)\beta_1(\theta,u_2-u_1)+ \cdots +(t-u_H)\beta_H(\theta,t-u_H) + x \right)}	\label{backlogrc} 
				\end{eqnarray}	
	\item Delay bound : The probabilistic bound on the delay in a network is given, for all $t \ge 0$, by
				\begin{eqnarray}
				P\left\{W(t) > d \right\}  &\le& \inf_{\theta > 0} \sum_{u_H=d}^{t} \cdots \sum_{u_2=d}^{u_3}\sum_{u_1=d}^{u_2} e^{\theta \left((t-d-u_1)\alpha(\theta,t-d-u_1) \right)} \cdot \nonumber\\
				& & \ \ \ e^{-\theta \left((u_2-u_1)\beta_1(\theta,u_2-u_1)+ \cdots +(t-u_H)\beta_H(\theta,t-u_H)\right)}	\label{delayrc}
				\end{eqnarray}
\end{enumerate}
\end{corollary}
To analyze the linear scaling of end-to-end probabilistic performance bounds from Corollary \ref{corollary:pbrc} we consider a tandem network of $H$ nodes connected in series as shown in Fig. \ref{fig:tandemnet}, with each node offering similar service characterized by the stochastic service process $S$ with the corresponding effective capacity $\beta$. Let the effective bandwidth $\alpha$ of the arrival traffic $A$ and the effective capacity $\beta$ of the service process $S$, satisfy the condition $\alpha(\theta,t) \le \alpha(\theta)$ and $\beta(\theta,t) \le \beta(\theta)$ for any $t,\theta \ge 0$. Then the probabilistic backlog bound from Corollary \ref{corollary:pbrc}, for any $\theta \ge 0$ and $t \rightarrow \infty$, will become
\begin{eqnarray}
P\left\{B(t) > x \right\} &\le&  \lim_{t \rightarrow \infty} \sum_{u_H=0}^{t} \cdots \sum_{u_2=0}^{u_3}\sum_{u_1=0}^{u_2} e^{\theta \left((t-u_1)\alpha(\theta) - (t-u_1)\beta(\theta) - x \right)}	\nonumber \\
&=& \frac{e^{-\theta x}}{\left(1-e^{-\theta(\beta(\theta) - \alpha(\theta))}\right)^H}\label{backlogr1}
\end{eqnarray}
The probabilistic delay bound from Corollary \ref{corollary:pbrc}, for any $\theta \ge 0$ and $t \rightarrow \infty$, will become
\begin{eqnarray}
P\left\{W(t) > d \right\} &\le&  \lim_{t \rightarrow \infty} \sum_{u_H=d}^{t} \cdots \sum_{u_2=d}^{u_3}\sum_{u_1=d}^{u_2} e^{\theta \left((t-d-u_1)\alpha(\theta) - (t-u_1)\beta(\theta) \right)}	\nonumber \\
&=& \frac{e^{-\theta \alpha(\theta)d}}{\left(1-e^{-\theta(\beta(\theta) - \alpha(\theta))}\right)^H} \label{delayr1}
\end{eqnarray}
We used the equality $\lim_{t \rightarrow \infty} \sum_{u_H=0}^{t} \cdots \sum_{u_2=0}^{u_3}\sum_{u_1=0}^{u_2} e^{-a(t-u_1)} = \frac{1}{(1-e^{-a})^H}$ for all $a \ge 0$ and the stability condition $\alpha(\theta) \le \beta(\theta)$ for all $\theta \ge 0$ in the equations (\ref{backlogr1}) and (\ref{delayr1}).

If the probabilistic backlog and delay bounds are violated at most with the probability $\varepsilon$, then setting the bounds on the right-hand sides of equations (\ref{backlogr1}) and (\ref{delayr1}) to $\varepsilon$ and solving for $x$ and $d$ gives
\begin{eqnarray}
x &\ge& -\frac{H}{\theta}\log{\left(1-e^{-\theta(\beta(\theta)-\alpha(\theta))}\right)} - \frac{\log{\varepsilon}}{\theta} \label{backlogr2}\\
d &\ge& -\frac{H}{\theta \alpha(\theta)}\log{\left(1-e^{-\theta(\beta(\theta)-\alpha(\theta))}\right)} - \frac{\log{\varepsilon}}{\theta  \alpha(\theta)} \label{delayr2}
\end{eqnarray}
It is apparent from equations (\ref{backlogr2}) and (\ref{delayr2}) that the end-to-end backlog and delay bounds using Corollary \ref{corollary:pbrc} grows linearly in the number of nodes $H$ a flow traverses in a network. 

The significance of the presented approach is that the stochastic information about the arrival and service processes are retained as long as possible using the concept of effective bandwidth and effective capacity, respectively, which allows the efficient computation of end-to-end stochastic performance measures than in other approaches as in \cite{yuming:2006,florin:2006,li:2007,jiang:2009-1} where the stochastic information about the arrival and service processes are lost as soon as statistical envelopes are fixed. 


\section{Numerical Example}
\label{sec:num}

\begin{figure}
\centering
\includegraphics[scale=0.5]{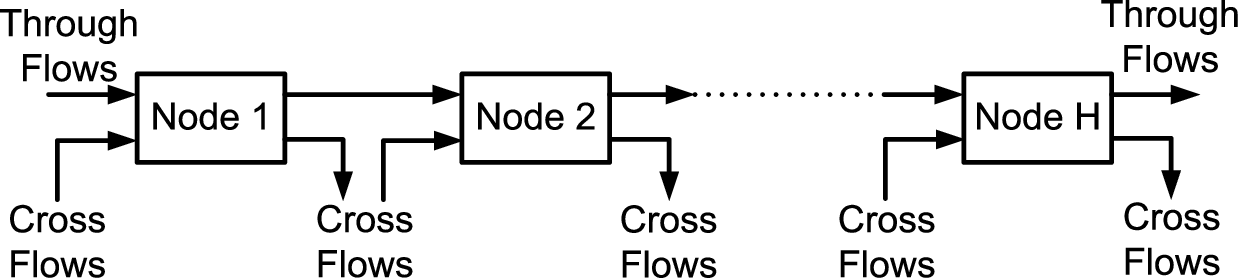}
\caption{ Network of H concatenated nodes with cross traffic}
\label{fig:tandemnet1} 
\vspace{-5 mm}
\end{figure}
In this section, we illustrate the benefits of end-to-end stochastic network calculus with effective bandwidth and effective capacity using the arrival traffic modelled as Markov modulated on-off (MMOO) process, especially we show that the end-to-end performance bounds computed using Theorem \ref{theorem:pbr} and Corollary \ref{corollary:pbrc} are as good as the ones obtain using statistical envelopes \cite{florin:2006}. Markov modulated on-off process is commonly used to model voice \cite{schwartz:1996} and video traffic \cite{maglaris:1988} in the Internet. Markov modulated on-off process can be in "`On"' state or "`Off"' state for random time intervals which are negative exponentially distributed with averages $E[T_{on}]$ and $E[T_{off}]$, respectively. In "`On"' state, arrival traffic transmits data at a constant rate $P$ and no data is transmitted in "`Off"' state. The effective bandwidth of Markov modulated on-off process has an interesting property that $\alpha(\theta,t) \le  \alpha(\theta)$ and for any $\theta > 0$ is given by 
\begin{equation}
\alpha(\theta) = \frac{1}{2\theta}\left(P\theta-r_{{10}}-r_{{01}}+\sqrt {\left(P\theta - r_{{10}} + r_{{01}} \right)^{2} + 4r_{{10}}r_{{01}} }\right)
\end{equation}
where $r_{10} = \frac{1}{E[T_{on}]}$ and $r_{01} = \frac{1}{E[T_{off}]}$.

For the analysis we consider a network of $H$ nodes connected in series with cross traffic as shown in Fig. \ref{fig:tandemnet1}. The queue at each hop $h$ is served in a work conserving fashion at a constant deterministic service rate $C$. The flow of interest is the one which traverses through the network of $H$ nodes connected in series and is termed through flow $A$. The flow which transits the network at each hop is termed cross flow $A_{c_h}$ for $h=1, \ldots , H$. Let $\alpha$ and $\alpha_{c_h}$ be the effective bandwidth functions of the through flow $A$ and the cross flow $A_{c_h}$, respectively, for $h=1, \ldots , H$. Let there be $N$ independent through flows at the ingress of the network and $M_h$ independent cross flows at each hop $h$ inside the network. The stability condition $C \ge N \alpha(\theta) + M_h \alpha_{c_h}(\theta)$ must be satisfied at each hop $h$, for $h=1, \ldots , H$ and any $\theta \ge 0$. 

The simplified network considered in Fig. \ref{fig:tandemnet1} can be seen as a section of a larger network where the arrival traffic traverses. No assumption is made about the topology of the larger network and can include both feedforward and networks with feedback traffic. However, two main assumptions were made about the considered network, firstly, stochastic information about the cross flow at each network node along the path of through flow is assumed to be known. Secondly, the traffic flow is assumed to follow the same path, i.e., routing is assumed to be fixed for the entire duration of its transmission.

To simplify the analysis, we assume all the cross flows have similar characteristics i.e., $A_{c_1} \equiv A_{c_2} \equiv \cdots \equiv A_{c_H} \equiv A_c$ \footnote{For any two random variables $X$ and $Y$, $X \equiv Y$ denotes that $X$ and $Y$ have the same distribution (the same cumulative distribution function (CDF))} and $M_1 = M_2 = \cdots = M_H = M$. The service available to the $N$ through flows at hop $h$ can be characterized using leftover stochastic service process $S_h(t) = S(t) = Ct - M A_{c}(t)$ with effective capacity function $\beta_h(\theta) = \beta(\theta) = C - M \alpha_c(\theta)$ under general scheduling model \cite{fidler:2006} for $h=1, \ldots, H$ and any $t, \theta \ge 0$. We evaluate the larger time interval [$0,\infty$] instead of [$0,t$] to compute end-to-end, closed-form performance measures using Theorem \ref{theorem:pbr} and Corollary \ref{corollary:pbrc}. The backlog $x$ and delay $d$ bounds which are violated at most with probability $\varepsilon$ can be computed from equations (\ref{backlogr}) and (\ref{delayr}), for any $t \ge 0$ and $\theta > 0$, as 
\begin{eqnarray}
x &\ge& -\frac{H(H+1)}{\theta}\log{\left(1-e^{-\theta( C - N \alpha(\theta) - M \alpha_c(\theta))}\right)} - \frac{H+1}{\theta}\log{\varepsilon} \label{backlogr3}\\
d &\ge& -\frac{H(H+1)}{\theta \alpha(\theta)}\log{\left(1-e^{-\theta( C - N \alpha(\theta) - M \alpha_c(\theta))}\right)} - \frac{H+1}{\theta  \alpha(\theta)}\log{\varepsilon} \label{delayr3}
\end{eqnarray}
For the case where arrival traffic and service offered at each network node are independent of one another, only feedforword network can be considered, and the backlog $x$ and delay $d$ bounds which are violated at most with probability $\varepsilon$ can be computed, for any $t \ge 0$ and $\theta > 0$, from equations (\ref{backlogr2}) and (\ref{delayr2}) 
\begin{eqnarray}
x &\ge& -\frac{H}{\theta}\log{\left(1-e^{-\theta(C - N \alpha(\theta) - M \alpha_c(\theta))}\right)} - \frac{\log{\varepsilon}}{\theta} \label{backlogr4}\\
d &\ge& -\frac{H}{\theta \alpha(\theta)}\log{\left(1-e^{-\theta(C - N \alpha(\theta) - M \alpha_c(\theta))}\right)} - \frac{\log{\varepsilon}}{\theta  \alpha(\theta)} \label{delayr4}
\end{eqnarray}
It is apparent from equations (\ref{backlogr4}) and (\ref{delayr4}) that the end-to-end backlog and delay bounds for MMOO traffic model using Corollary \ref{corollary:pbrc} grows linearly in the number of nodes $H$ a flow traverses in a network, under the assumption of independence.
\begin{figure}
\centering
\includegraphics[angle=-90,scale=0.4]{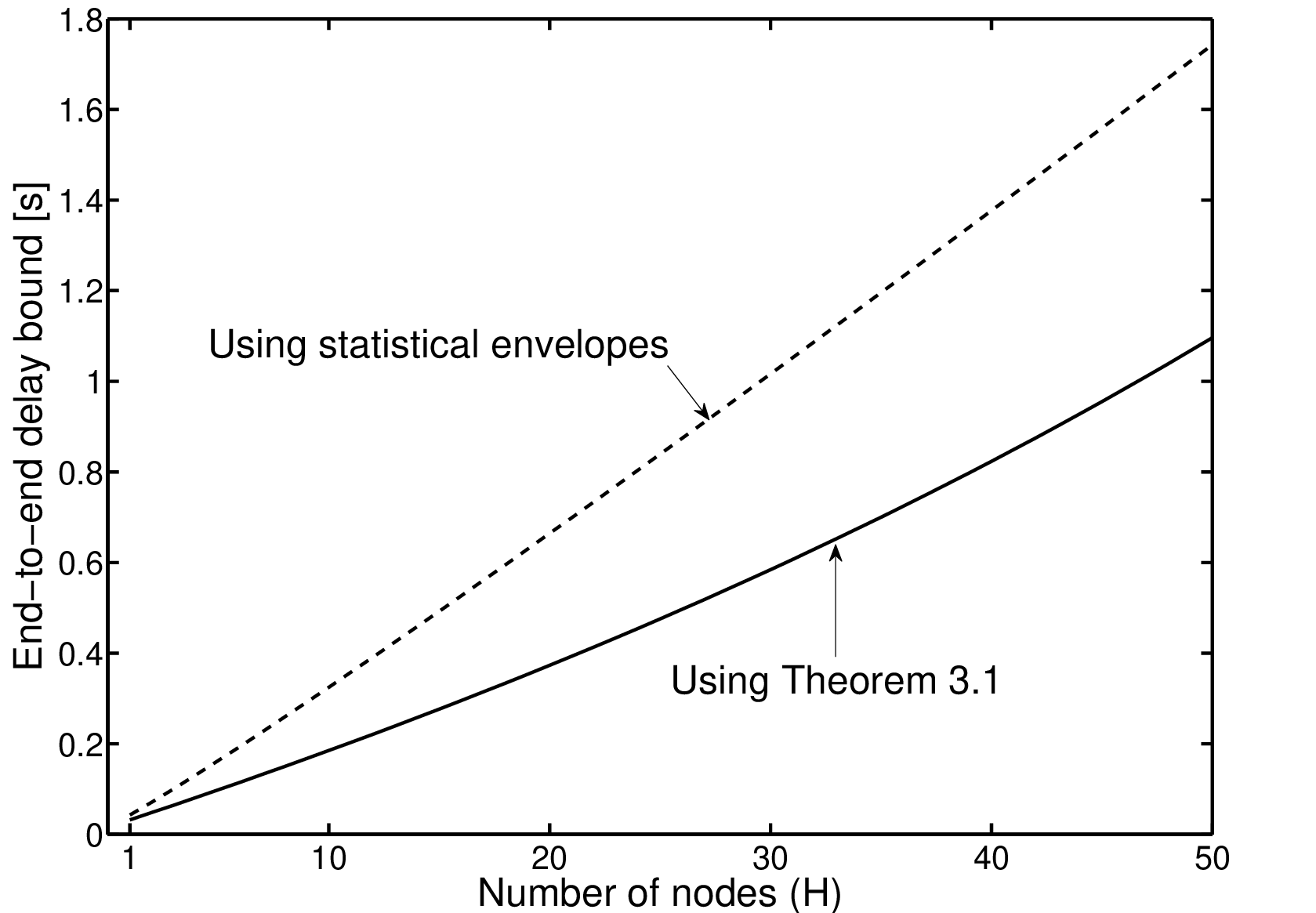}
\caption{End-to-end delay bound with violation probability $\varepsilon = 10^{-9}$ for Markov modulated on-off traffic in a network of increasing number of nodes $H$ with $N = 781$ through flows and $M = 1953$ cross flows at each hop}
\label{fig:scaleplot} 
\vspace{-5 mm}
\end{figure}
\begin{figure}
\centering
\includegraphics[angle=-90,scale=0.4]{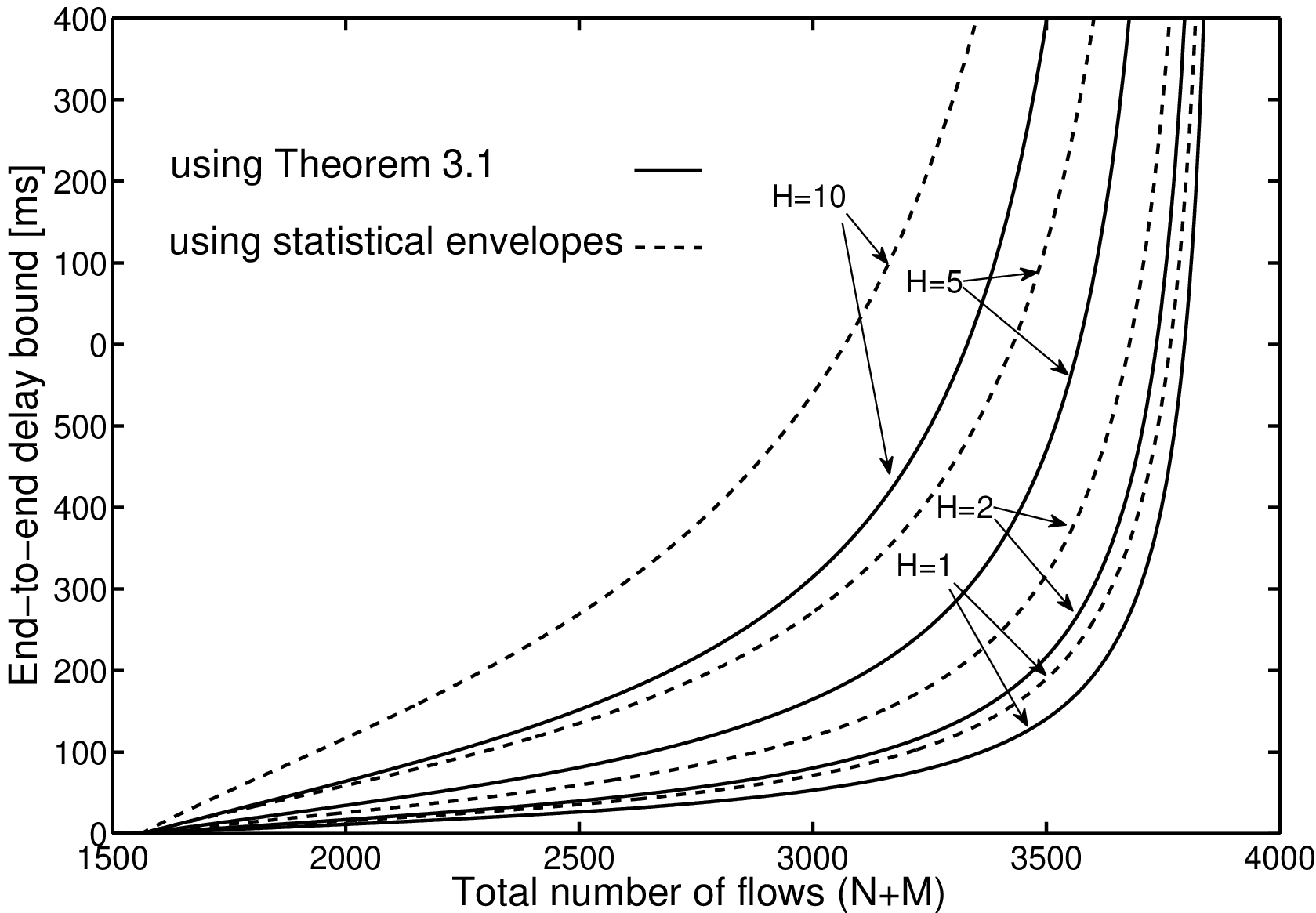}
\caption{End-to-end statistical delay bounds for Markov modulated on-off traffic in a network of $H$ nodes with violation probability $\varepsilon = 10^{-9}$}
\label{fig:delayplot} 
\vspace{-5 mm}
\end{figure}
\begin{figure}
\centering
\includegraphics[angle=-90,scale=0.4]{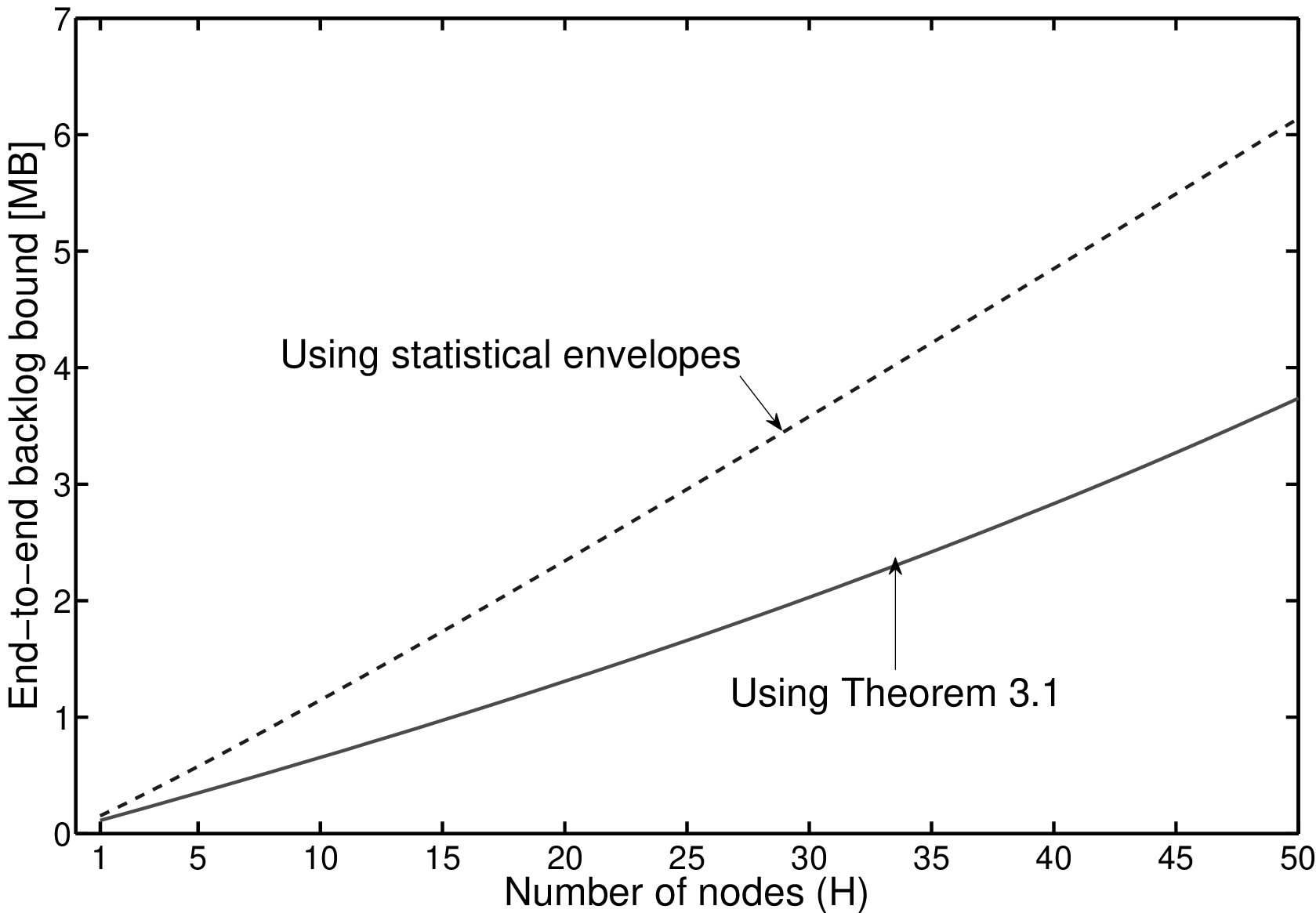}
\caption{End-to-end backlog bound with violation probability $\varepsilon = 10^{-9}$ for Markov modulated on-off traffic in a network of increasing number of nodes $H$ with $N = 781$ through flows and $M = 1953$ cross flows at each hop}
\label{fig:scaleplot1} 
\vspace{-5 mm}
\end{figure}
\begin{figure}
\centering
\includegraphics[angle=-90,scale=0.4]{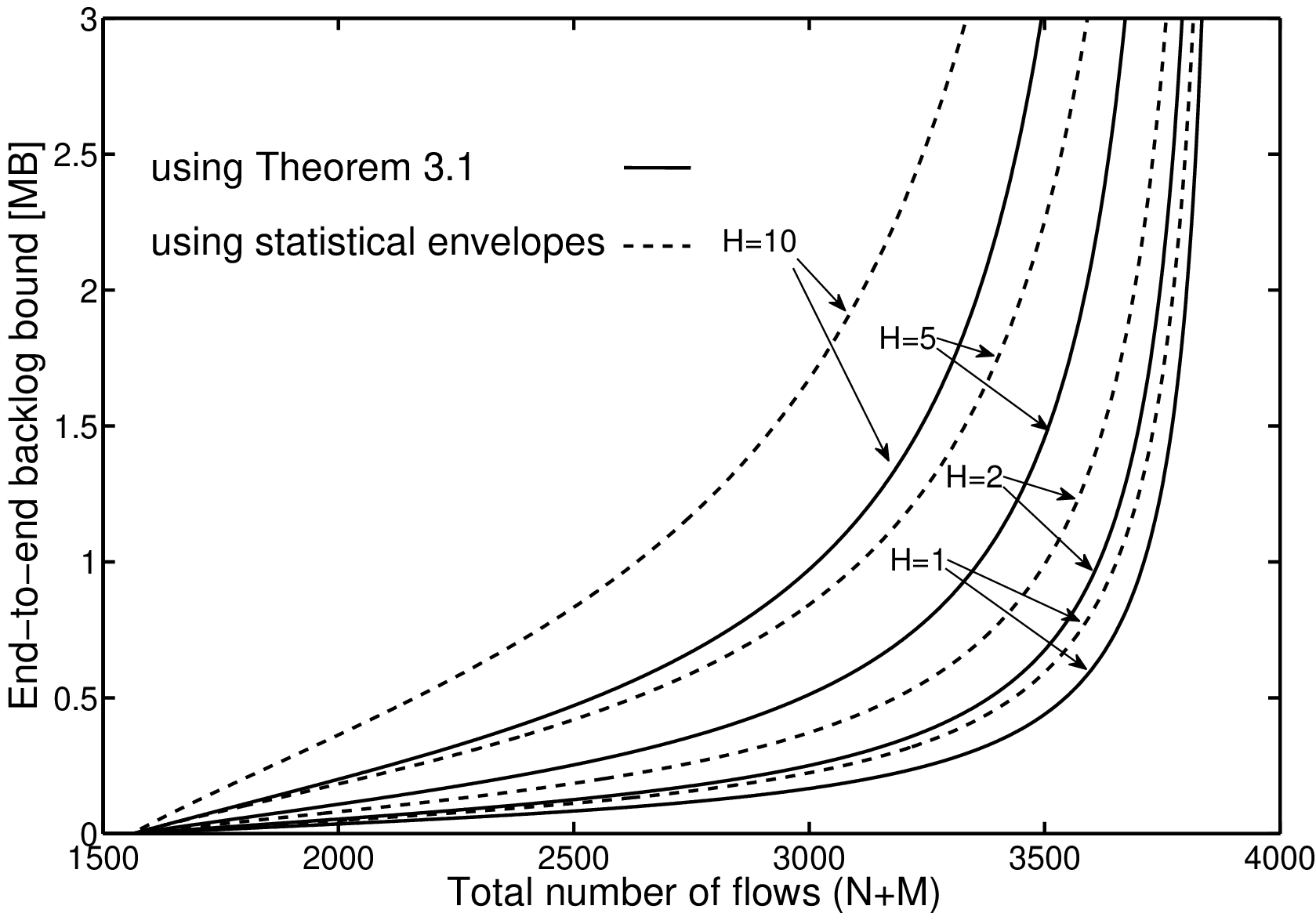}
\caption{End-to-end statistical backlog bounds for Markov modulated on-off traffic in a network of $H$ nodes with violation probability $\varepsilon = 10^{-9}$}
\label{fig:delayplot1} 
\vspace{-5 mm}
\end{figure}
For the numerical experiment, we compute the end-to-end delay and backlog bound for $N$ through flows in the network  with a violation probability $\varepsilon = 10^{-9}$. The capacity of the server $C$ at each hop is set to $100Mbps$. To simplify the numerical analysis, all arrival traffic at each node in the network is assumed to be Markov modulated on-off process (MMOO) with similar characteristics, i.e., $A \equiv A_c$. However, through flows and cross flows can be of different traffic types and will have no influence on the analysis as long as effective bandwidth functions for the considered arrival traffic exists. For each Markov modulated on-off  process, we choose the following values which are typically used to model voice flows \cite{schwartz:1996}: $P=64Kbps$, $E[T_{on}]=0.4s$ and $E[T_{off}]=0.6s$. The average arrival rate ($m$) of the Markov modulated on-off traffic using the given parameters is $25.6Kbps$. 

\begin{figure}
\centering
\includegraphics[angle=-90,scale=0.4]{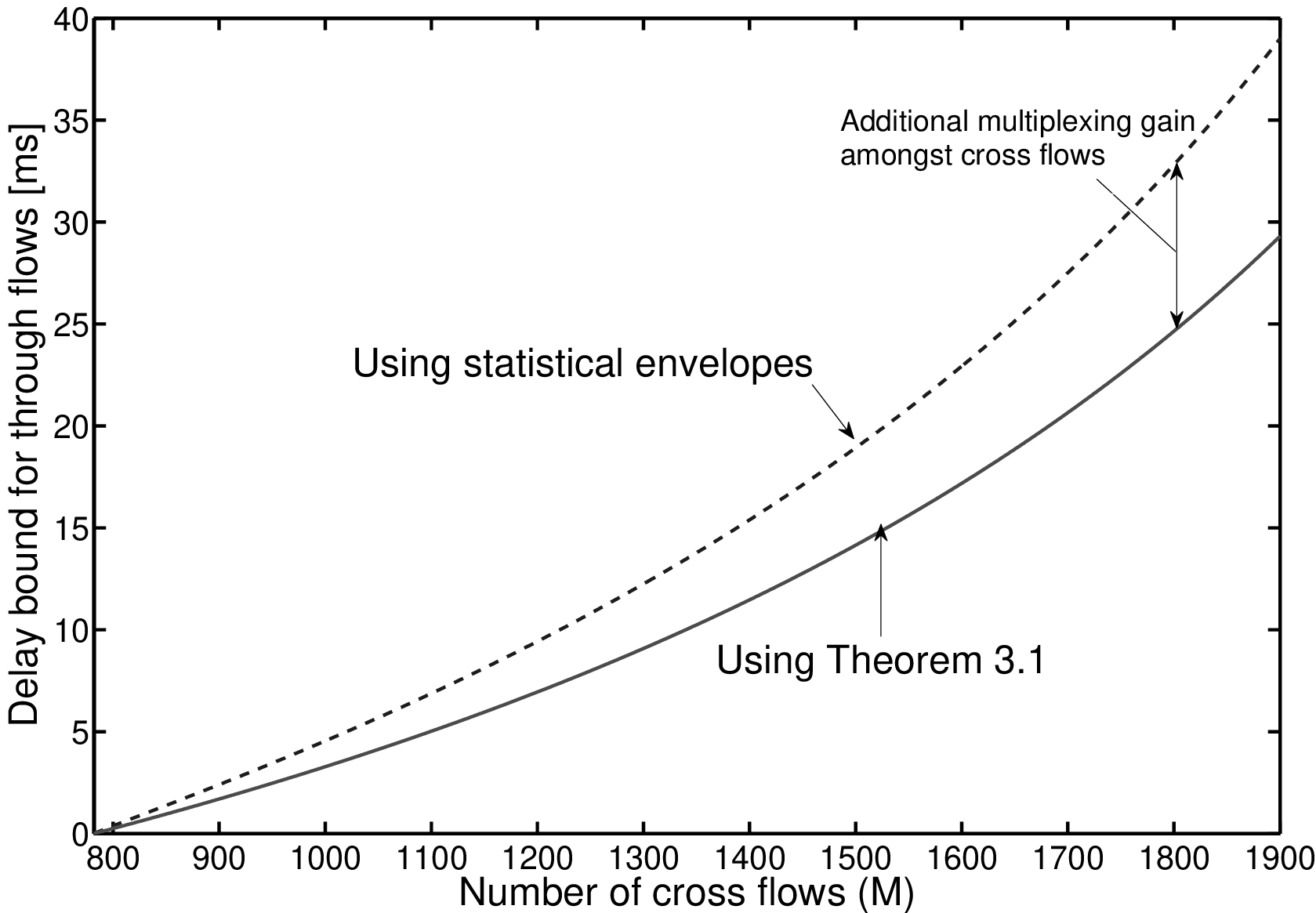}
\caption{Impact of cross flows on the delay bound of through flows}
\label{fig:crossplot} 
\vspace{-5 mm}
\end{figure}

In Figs. \ref{fig:scaleplot} and  \ref{fig:delayplot}, the end-to-end delay bound computed using statistical envelope definitions from \cite{florin:2006} and delay bound from equation (\ref{delayr3}) are plotted. In Figs. \ref{fig:scaleplot1} and  \ref{fig:delayplot1}, we plot the end-to-end backlog bound computed using statistical envelope definitions from \cite{florin:2006} and backlog bound from equation (\ref{backlogr3}). Fig. \ref{fig:scaleplot} and Fig. \ref{fig:scaleplot1} shows the probabilistic end-to-end delay and backlog bound with a violation probability $(\varepsilon)$ of $10^{-9}$ as a function of increasing number of hops $H$. At each hop, $N = 781$ through flows are multiplexed with $M = 1953$ independent cross flows. In Fig. \ref{fig:delayplot} and Fig. \ref{fig:delayplot1}, we plot the probabilistic end-to-end delay and backlog bounds for $N$ through flows in a network with $H=1,2,5,10$ hops for increasing $N+M$ number of flows at each hop while maintaining $N=M$. It can be observed that the delay and backlog bounds from equations (\ref{delayr3}) and (\ref{backlogr3}), respectively, yield a tighter bounds than the ones computed using statistical envelopes \cite{florin:2006}. The tighter bounds are achieved using the new approach as the stochastic information about the arrival and service processes are retained as long as possible using the concept of effective bandwidth and effective capacity, respectively, which allows the efficient computation of end-to-end stochastic delay and backlog bonds than in other approaches as in \cite{yuming:2006,florin:2006,li:2007,jiang:2009-1} where the stochastic information about the arrival and service processes are lost as soon as statistical envelopes are fixed. Fig. \ref{fig:crossplot} shows the impact of cross flows on the delay bound for the through flows at single node. The number of through flows ($N$) is set at $781$, to determine the delay bound for  through flows as the number of cross flows ($M$) is increased from $781$ to $1950$. It can be observed from the Fig. \ref{fig:crossplot} that the use of effective bandwidth and effective capacity functions of arrival and service processes, respectively, allows to capture additional statistical multiplexing gain amongst independent cross traffic in comparison to use of statistical envelopes. This has a direct influence on the delay bounds for through flows at a network node as we modelled the service available to the through flows from the left over service after serving the cross flows in the network node.
\section{Conclusion}
\label{sec:conclusion}
We presented an end-to-end stochastic network calculus with effective bandwidth and effective capacity functions. We then showed that such a formulation of network calculus results in end-to-end performance measures that grow linearly in the number of nodes traversed by the arrival traffic, under the assumption of independence. Further we showed using numerical example with Markov modulated on-off arrivals traffic that the end-to-end delay and backlog bound computed using the presented stochastic network calculus are tighter than the bounds obtained using statistical envelopes.





\bibliographystyle{elsarticle-num}
\bibliography{biblio}







\end{document}